 \documentclass[pmlr,twocolumn,10pt]{jmlr} 

\usepackage{booktabs}

\usepackage{multirow}
\usepackage{graphicx}
\usepackage{xcolor}
\usepackage{colortbl}
\usepackage{cleveref}
\usepackage{hyperref}

\usepackage[switch]{lineno}

\theorembodyfont{\upshape}
\theoremheaderfont{\scshape}
\theorempostheader{:}
\theoremsep{\newline}

\jmlrvolume{LEAVE UNSET}
\jmlryear{2024}
\jmlrsubmitted{LEAVE UNSET}
\jmlrpublished{LEAVE UNSET}
\jmlrworkshop{Machine Learning for Health (ML4H) 2024} 

 \title[New Test-Time Scenario]{New Test-Time Scenario for Biosignal: Concept and Its Approach}

\author{%
\Name{Yong-Yeon Jo} \Email{yy.jo@medicalai.com} 
\addr MedicalAI Co. Ltd.
\AND
\Name{Byeong Tak Lee} \Email{bytaklee@medicalai.com} 
\addr MedicalAI Co. Ltd.
\AND
\Name{Beom Joon Kim} \Email{kim.bj.stroke@gmail.com}
\addr Seoul National University Bundang Hospital
\AND
\Name{Jeong-Ho Hong} \Email{neurohong79@gmail.com}
\addr Keimyung University Dongsan Medical Center
\AND
\Name{Hak Seung Lee} \Email{cardiolee@medicalai.com}
\addr MedicalAI Co. Ltd.
\AND
\Name{Joon-myoung Kwon} \Email{cto@medicalai.com}
\addr MedicalAI Co. Ltd.
}

\begin{document}

\maketitle

\begin{abstract}
Online Test-Time Adaptation (OTTA) enhances model robustness by updating pre-trained models with unlabeled data during testing. In healthcare, OTTA is vital for real-time tasks like predicting blood pressure from biosignals, which demand continuous adaptation. We introduce a new test-time scenario with streams of unlabeled samples and occasional labeled samples. Our framework combines supervised and self-supervised learning, employing a dual-queue buffer and weighted batch sampling to balance data types. Experiments show improved accuracy and adaptability under real-world conditions.

\end{abstract}
\begin{keywords}
Test Time Adaptation (TTA), Biosignals, Electrocardiogram (ECG), Photoplethysmogram (PPG)
\end{keywords}

\section{Introduction}
Online Test-Time Adaptation (OTTA) enhances model robustness to distribution shifts by continuously updating pre-trained models with unlabeled data during testing (\cite{chen2023improved,gan2023decorate,pmlr-v119-sun20b}). 
Such OTTA is a crucial approach for predictive tasks in real-time monitoring and intervention within the healthcare domain.
These tasks rely on various biosignals, including electrocardiograms (ECG) and photoplethysmography (PPG), which exhibit unique patterns that dynamically change over time, reflecting an individual’s health status (\cite{vollmer2022simultaneous, gow2023mimic}). 
This dynamic nature makes it challenging to maintain prediction accuracy with static models. 
In practice, to address this, models often require ongoing calibration through periodic manual measurements, which previous studies have not fully addressed.

We introduce a new test-time paradigm for biosignal applications, enabling models to adapt using continuous streams of unlabeled data and occasional labeled samples that simulate periodic calibration. To address this, we propose a dual-head framework (\cite{pmlr-v119-sun20b}) supported by techniques like a dual-queue buffer and weighted batch sampling to balance learning from both data types. This approach improves adaptability and prediction accuracy in dynamic healthcare environments.

\begin{figure*}[h!]
    \centering
    \includegraphics[width=0.9\textwidth]{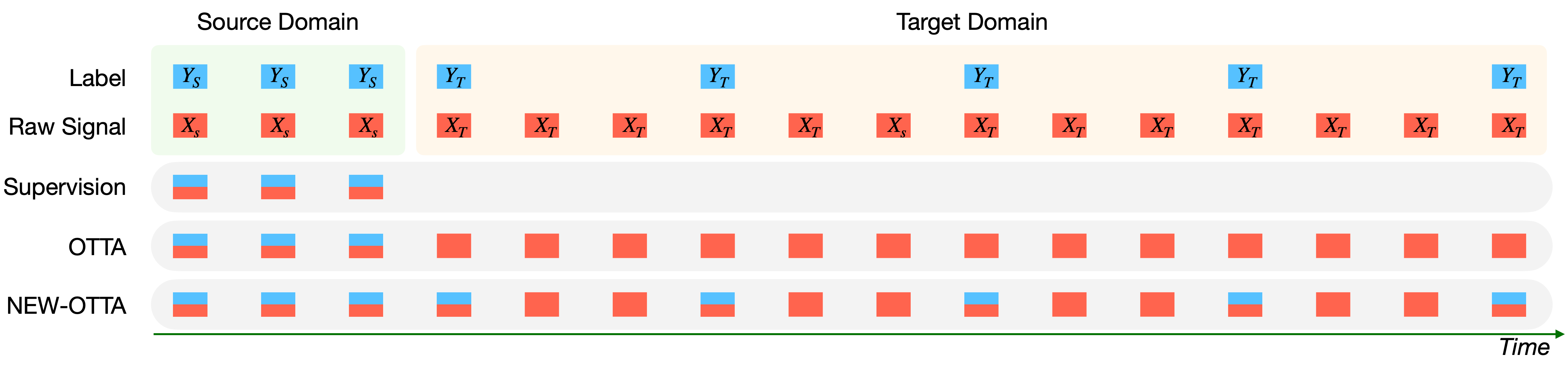} 
\vspace{-3mm}
    \caption{
    Example of data flow and approaches for new test-time paradigm.
    } 
    \label{fig:new-otta}
\end{figure*}

To implement the proposed test scenario, we utilize the PulseDB dataset (\cite{wang2023pulsedb}), a benchmark for cuff-less blood pressure estimation. Our evaluation adjusts the frequency of label injection during testing to simulate varying conditions. Additionally, as labeled samples accumulate, the model is fine-tuned periodically, with predictions temporarily paused for updates. This setup enables a thorough assessment of the model’s adaptability and robustness under different levels of label availability and injection frequencies. 
Our experiments demonstrate that frequent label injections and larger initial labeled datasets improve model adaptability and accuracy in blood pressure prediction, with ablation studies highlighting the buffer size and sampling strategy's trade-offs.


\section{New Test-Time Scenario}

Supervised models are typically trained assuming source domain availability, without accounting for data influx during testing. However, real-world scenarios often involve distribution shifts between training (source) and deployment (target) data due to factors like environmental changes, sensor variability, or user behavior. OTTA addresses this by updating models with unlabeled target domain samples before inference, enabling adaptation to real-time changes (\cite{wang2020tent, niu2022eata, wang2022continual}).

In healthcare, biosignals such as ECG and PPG dynamically reflect health status and can shift with sudden condition changes. Maintaining prediction accuracy requires regular calibration using manually collected measurements as labels.
Real-world examples include blood pressure monitoring (\cite{samsung_health_monitor}), ICU patient monitoring (\cite{urden2013critical}), and continuous glucose monitoring (\cite{galindo2020continuous}).

We introduce a new test-time scenario tailored for biosignal applications, as illustrated in Figure~\ref{fig:new-otta}. In this scenario, raw signals are continuously available from both source and target domains, while labels are consistently available in the source domain but sporadically available in the target domain.

Supervised models rely solely on labeled source domain data for training. OTTA approaches utilize raw signals from both domains but only labels from the source domain. The proposed new OTTA paradigm leverages all available data, including raw signals and any intermittent labels from both domains, during both training and testing phases.

This scenario addresses the challenge of adapting models trained on labeled source domain data to target domain data with distribution shifts, using a mix of continuous unlabeled data and intermittent labeled samples.

\begin{figure*}[h!]
    \centering
    \includegraphics[width=0.9\textwidth]{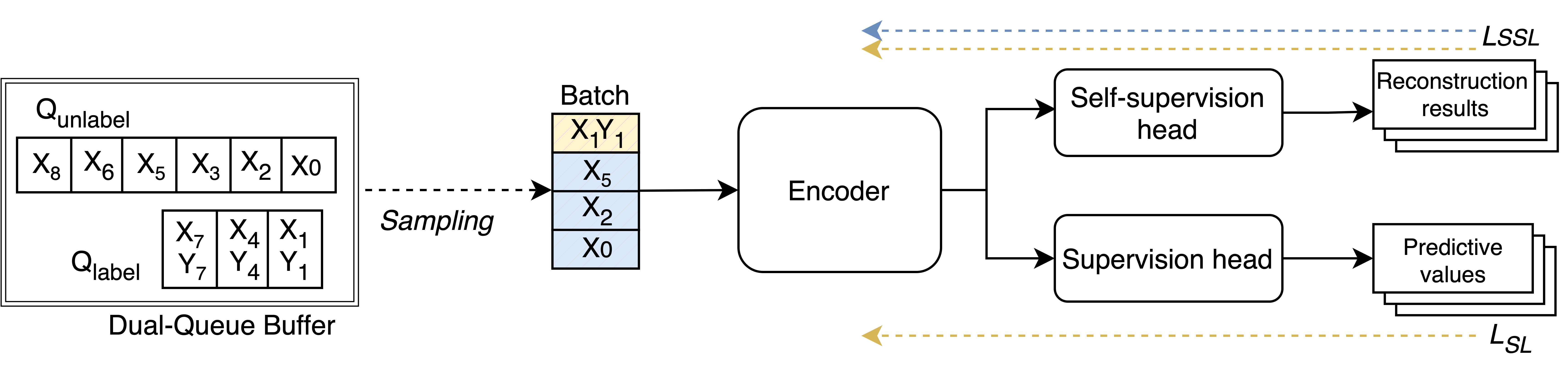} 
    \caption{Proposed OTTA framework.} 
    \label{fig:method}
\end{figure*}

\section{Proposed Method}
We propose a framework for a new test-time paradigm in biosignal applications. Our OTTA framework (Figure~\ref{fig:method}) operates during the test phase by using a model that simultaneously handles both supervised learning (SL) and self-supervised learning (SSL) tasks, enabling efficient adaptation to real-time labeled and unlabeled samples.

Previous studies like Liu et al.\ (\cite{liu2021ttt++}) introduced a dual-head approach with separate SL and SSL heads. While the model learns from both tasks during training, it relies only on the SSL head during testing to adapt to unlabeled data.

Our approach dynamically adapts model parameters to optimize performance on both unlabeled and scarce labeled samples during testing, involving both calibration and adaptation by accommodating both sample types. Notably, any model capable of handling SL and SSL tasks can be used in our framework.

While OTTA methods typically use batch-based learning (\cite{wang2020tent, chen2022adacontrast, niu2022eata}), treating all incoming labeled and unlabeled samples uniformly can hinder effective use of labeled samples. Despite their infrequency, labeled samples provide crucial information but can be overshadowed due to scarcity.

To handle dynamic data influx, we introduce a dual-queue buffer system with separate queues for unlabeled and labeled samples. The unlabeled queue is larger due to more frequent data, while the labeled queue is smaller. This prevents rapid depletion of labeled data, ensuring the model benefits from their crucial information. The influx of unlabeled samples keeps the model updated with recent trends, and expanding the buffer size allows learning from a broader range of past data, enhancing recognition of long-term patterns.

We introduce a weighted batch sampling strategy to effectively leverage the dual-queue buffer, sampling from Queues in specific proportions. A higher proportion of unlabeled samples helps capture recent trends but may lack precision due to the absence of labels; too many labeled samples risk overfitting on redundant data. We balance sampling to optimize adaptability and accuracy.

By mixing these samples, the model continuously learns from recent data inputs while maintaining accuracy through the effective use of labeled data. This ensures that the model adapts to new patterns while retaining the valuable information provided by the labeled samples.

\section{Experiments}

\subsection{Experimental Setup}


\paragraph{Data:} We selected the PulseDB dataset specifically designed for benchmarking cuff-less blood pressure estimation~(\cite{wang2023pulsedb}). 
This dataset is constructed from the MIMIC-III waveform~(\cite{johnson2016mimic}) and the VitalDB databases~(\cite{lee2022vitaldb}), and it contains over 5.2 million high-quality 10-second segments of electrocardiogram (ECG), photoplethysmogram (PPG), and arterial blood pressure (ABP) waveforms from over 5k subjects, along with demographic information.

\paragraph{Evaluation:} We established diverse evaluation settings to simulate real-world scenarios by varying the frequency of labeled target samples, ranging from no labels available to frequent label injections. This included scenarios where the model is fine-tuned with up to 50 initial labeled samples, simulating preliminary biosignal measurements before application use. For evaluation metrics, we used mean absolute error (MAE) and the correlation between predicted and actual values for systolic and diastolic blood pressure (SBP and DBP). To ensure fair comparisons, we fixed the number of test samples used for inference and excluded the initial labeled samples and those used for label injections from performance evaluations. As a result, any samples affected by the label injection frequency were not considered in the evaluation.

\vspace{-3mm}
\begin{table*}[ht]
\small
\caption{Results of blood pressure prediction depending on the label injection frequency and the number of initial labels available before starting the test-time adaptation}
\label{tab:result}
\centering
\begin{tabular}{c|c|cc|cc|cc|cc}
\toprule
\multirow{2}{*}{\parbox{2cm}{\centering \textbf{Injection}\\ \textbf{Frequency}}} & \textbf{\# of Init. Labels} & \multicolumn{2}{c}{\textbf{0}} & \multicolumn{2}{|c}{\textbf{10}} & \multicolumn{2}{|c}{\textbf{20}} & \multicolumn{2}{|c}{\textbf{50}} \\ \cmidrule(lr){2-10}
 & & \textbf{SBP} & \textbf{DBP} & \textbf{SBP} & \textbf{DBP} & \textbf{SBP} & \textbf{DBP} & \textbf{SBP} & \textbf{DBP} \\ \midrule
 
 \multirow{2}{*}{\textbf{N/A}} & \textbf{MAE} & \cellcolor{red!30} 13.56 & \cellcolor{red!30} 8.12 & 14.74 & 8.91 & 14.32 & 8.59 & \cellcolor{green!30} 13.58 & \cellcolor{green!30} 8.17 \\ \cmidrule(lr){2-10} 
 & \textbf{Correlation} & \cellcolor{red!30}0.55 & \cellcolor{red!30}0.50 & 0.44 & 0.37 & 0.48 & 0.43 & \cellcolor{green!30} 0.55 & \cellcolor{green!30} 0.51 \\ \midrule
 
 \multirow{2}{*}{\textbf{100}} & \textbf{MAE} & 14.25 & 8.64 & 13.87 & 8.37 & 13.62 & 8.15 & 13.00 & 7.82 \\ \cmidrule(lr){2-10} 
 & \textbf{Correlation} & 0.44 & 0.39 & 0.53 & 0.48 & 0.56 & 0.53 & 0.61 & 0.58 \\ \midrule
 
 \multirow{2}{*}{\textbf{50}} & \textbf{MAE} & 13.42 & 8.13 & 13.24 & 7.95 & 13.02 & 7.78 & 12.69 & 7.64 \\ \cmidrule(lr){2-10} 
 & \textbf{Correlation} & 0.55 & 0.50 & 0.59 & 0.57 & 0.62 & 0.59 & 0.63 & 0.61 \\ \midrule
 
 \multirow{2}{*}{\textbf{20}} & \textbf{MAE} & 12.13 & 7.35	& 11.94 & 7.06 & 12.00 & 7.17	& 11.67 & 6.99   \\ \cmidrule(lr){2-10} 
 & \textbf{Correlation} & 0.68 & 0.67 & 0.70 & 0.70 & 0.70 & 0.69 & 0.72 & 0.71 \\ \midrule
 
\multirow{2}{*}{\textbf{10}} & \textbf{MAE} & \cellcolor{blue!30}10.98 & \cellcolor{blue!30}6.59 & 10.85 & 6.44 & 10.98 & 6.59 & 10.72 & 6.42 \\ \cmidrule(lr){2-10} 
 & \textbf{Correlation} & \cellcolor{blue!30}0.77 & \cellcolor{blue!30}0.76 & 0.78 & 0.78 & 0.77 & 0.76 & 0.78 & 0.77 \\ \bottomrule
\multicolumn{10}{l}{Model performance without any adaptation: [MAE] 13.66/8.15; [Correlation] 0.54/0.54 (SBP/DBP).} \\ 
\end{tabular}
\end{table*}


\paragraph{Model:} We adopted the Test-Time Training Online (TTT) framework (\cite{gandelsman2022test}). Our network consists of an encoder, a decoder as the self-supervision head, and a regressor as the supervision head. We used the Vision Transformer (ViT) (\cite{dosovitskiy2020image}) as the backbone encoder. The decoder performs signal reconstruction on masked signals, and the regressor predicts blood pressure (BP) values.

\paragraph{Training setup:}
For constructing the pre-trained model, we followed the TTT protocol outlined in \cite{gandelsman2022test}. Each sample was split into segments equivalent to tokens used in ViT. The model was updated sequentially: first through a reconstruction task (self-supervised learning, SSL) using mean squared error, followed by a prediction task (supervised learning, SL) using Shrinkage Loss \cite{lu2018deep}. We utilized the LAMB optimizer \cite{you2019large} and employed Population Based Training \cite{jaderberg2017population} to tune hyperparameters such as learning rate and weight decay, setting different values for each task. 

\paragraph{Test setup:}
We implemented and utilized the dual-queue buffer and sampling strategy exclusively during the test phase. The buffer consists of two queues,  $Q_{\text{unlabel}}$ and  $Q_{\text{label}}$, with sizes in a 2:1 ratio. In the sampling strategy, we selected samples in a 3:1 ratio within a total batch size of 32. The input settings were kept consistent with the training setup. Following the adaptation logic from \cite{gandelsman2022test}, the model was updated using the loss from the reconstruction task when presented with unlabeled samples. When labeled samples were available, the model was updated using a combined loss from both reconstruction and prediction tasks. For model optimization, we employed Stochastic Gradient Descent (SGD). This process was conducted individually for each subject, and the model update was repeated 10 times per batch.

\subsection{Experimental Results}

Table~\ref{tab:result} presents the mean absolute error (MAE) and correlation results for systolic blood pressure (SBP) and diastolic blood pressure (DBP) across various scenarios. The red cells represent the performance of the typical TTA approach, the green cells show the performance of the fine-tuned model using available initial labeled samples, and the blue cells indicate the performance achieved solely by our proposed approach. Compared to the baseline performance of the supervision model (MAE values of 13.66 for SBP and 8.15 for DBP, and correlation coefficients of 0.54 for both), most methods in the table show improved performance, indicating that TTA approaches enhance the model’s adaptability.

Increasing the label injection frequency generally decreases the MAE and improves the correlation between predicted and actual values, demonstrating better model adaptation to the target domain. However, beyond a certain point, additional label injections do not yield significant performance gains, suggesting a limit to their benefit. Similarly, increasing the number of initial labeled samples before starting the TTA task generally improves performance, as the model is better fine-tuned to the target domain’s data distribution. Yet, in some cases, the performance plateaus or shows only marginal improvement, indicating that beyond a certain threshold, more initial labeled samples may not provide significant advantages. Overall, our proposed framework outperforms the conventional TTA approach by effectively utilizing newly injected labeled data and adequately preparing the model with initial labeled samples.

\section{Conclusion}

We presented a test-time paradigm for biosignal applications, where models adapt to continuous streams of unlabeled data with occasional labeled samples. We propose a model handling supervised and self-supervised learning, supported by a dual-queue buffer and weighted batch sampling to utilize both data types effectively. Our evaluation using the PulseDB dataset shows that increasing label injection frequency and initial labeled samples improves blood pressure prediction, demonstrating adaptability and robustness. These findings highlight the potential to enhance real-time monitoring and prediction, creating a foundation for more resilient models capable of handling distribution shifts.

\vspace{-3mm}
\acks{This work was supported by the National Research Foundation of Korea (NRF) grant funded by the Korea government (Ministry of Science and ICT) (No. 2022R1A2C1009047).}

\newpage


\bibliography{jmlr-sample}






\end{document}